# Generation of Streaming Beam-Plasma Instability in Variable Lunar Plasma around Moon

Vipin K. Yadav*[1], Mahima Agarwal [2], Mehul Chakraborty [3], and Rajneesh Kumar [2]

(1) Space Physics Laboratory (SPL), Vikram Sarabhai Space Centre (VSSC), Thiruvananthapuram 695022, India
(2) Department of Physics, Banaras Hindu University (BHU), Varanasi 221005, India
(3) Laboratoire de Physique et Chimie de l'Environnement et de l'Espace, Centre National de la Recherche Scientifique (CNRS), Universite d'Orleans, Orleans 45071, France
* Email: vipin_ky@vssc.gov.in

## Abstract

Two-stream instability (TSI) is studied analytically in the lunar plasma environment. The electrons in the solar wind constitute the electron-beam and the lunar electron plasma constitutes the background plasma with which the electron-beam interacts to trigger the TSI. The lunar plasma is considered to have a variable proportion of the energetic (hot) electrons, 1% to 25% of the total lunar electrons, along with the bulk thermal (cold) population. The analysis shows that the presence of energetic electrons in the lunar plasma environment modify the TSI dispersion relation and can have a significant impact on the triggering of TSI and are capable of triggering non-linear phenomena by making the lunar plasma system unstable.

## 1. Introduction

The phenomena of plasma instability generation can occur anytime in space at any location where the local plasma environment is suitable for its sustenance. In general, plasma instability is a process in which a non-linear energy build-up takes place in a plasma system due to the availability of free energy source which drives the system unstable so that it is unable to hold the excess energy. Then the plasma system discards this excess energy either in the form of throwing away the particles which carries sufficient energy of in the form of plasma waves so that the plasma system regains its equilibrium energy configuration.

The relative motion between two plasma systems can trigger the streaming plasma instabilities. Generally, in a streaming plasma instability, there are two counter-moving plasma beams or a plasma beam passing through another plasma which is either moving very slowly with respect to the passing beam or is relatively at rest. The examples of various types of streaming plasma instabilities are – Ion-acoustic instability, Electron cyclotron drift instability, Ion-ion streaming instability, Lower-hybrid drift instability, two-stream instability, modified two stream instability, electron two-stream instability, etc. The streaming plasma instabilities are observed and studied at locations in the heliosphere.

The paper is organized as follows: In section 2, the TSI around the Moon is discussed. Section 3 describes the modified dispersion relation for the lunar TSI. In section 4, the lunar TSI (LTSI) dispersion relation is analyzed where the various hot electron proportions are considered. Section 5 contains the main conclusions drawn from this study.

## 2. Two-stream Instability around Moon

In the earlier study the lunar two-stream instability generation is analyzed considering only the non-energetic electrons participating in the interaction with the incoming solar wind [1, 2].

In this analysis, an electron beam, which is streaming through a plasma where the dominant species is electrons but their energy is lower than the incoming electron beam in such a way that $v_0 \approx 0$. It is further assumed in this analysis that

1. The background plasma is sufficiently large to interact with the incoming solar wind and is homogeneous and isotropic for the plasma parameters for both electrons and ions.
2. The ions in background plasma form a uniform positively charged background with no significant individual movement as the mobility of ions is many orders less than that of electrons. Hence, in this background plasma, the mobility is due to electrons only.
3. This is an electrostatic system with no magnetic field present around it. This treatment is just analytic, what exactly happens to the TSI can only be fully understood with simulation visualization. This analysis only indicates that there will be change but in what way requires simulations to be carried out.
4. The plasma oscillations are longitudinal and electrostatic in nature.
5. The electron number density in the incoming beam is much less than the background electron plasma.

The lunar electron energy distribution suggests the maximum electron energy in the lunar ionosphere to be ≤ 5 eV [3].

In recent times, the observations by instruments onboard Acceleration, Reconnection, Turbulence and Electrodynamics of the Moon's Interaction with the Sun (ARTEMIS) spacecraft showed that the average electron temperature in lunar ionosphere is ≈ 2.2 eV [4].

The average electron temperature in the solar wind is ≈ 10 eV [5] in a typical range of 7 - 17 eV. On an extended range, the solar wind electron temperature can be taken as 5 - 30 eV [6]. These observations lead us to conclude that even on a quiet solar day, when no solar activity is taking place, the maximum lunar electron energy is less than the impinging solar wind electrons.

For this analysis also, the above assumptions are considered valid and along with that the beam plasma is considered as magnetohydrodynamic (MHD) fluid. There are three main parameters which are to be considered - the electron number density, the electron velocity and the electric field for which the following three equations are used.

The electron momentum equation is considered to establish a relation between the electron velocity ($v_e$) and the electric field E which is given by

$$m_e n_e \left[\frac{\partial v_e}{\partial t} + \vec{v_e}.\nabla v_e\right] = -en_e\vec{E} \qquad (1)$$

The continuity equation for electrons is considered to link the electron number density ne and the electron velocity $v_e$ which is given by

$$\frac{\partial n_e}{\partial t} + \vec{\nabla}.(n_e v_e) = 0 \qquad (2)$$

To complete the analysis, Gauss law is taken to establish a relation between the electric field $\vec{E}$ and the electron density $n_e$ which is given by

$$\vec{\nabla}.\vec{E} = \frac{\rho}{\epsilon_0} \qquad (3)$$
$$\epsilon_0 \vec{\nabla}.\vec{E} = \rho = e(n_i - n_e) \qquad (4)$$

All other symbols/variables have their usual standard meanings. These equations are linearized with

$$n_e = n_{b0} + \widetilde{n_b} \qquad (5)$$
$$\vec{v_e} = \vec{v_{b0}} + \widetilde{\vec{v_b}} \qquad (6)$$

After linearizing, the electron momentum equation becomes

$$\frac{\partial \widetilde{\vec{v_b}}}{\partial t} + \vec{v_{b0}}.\vec{\nabla}\widetilde{\vec{v_b}} = -\frac{e\vec{E}}{m_e} \qquad (7)$$

the continuity equation becomes

$$\frac{\partial \widetilde{n_b}}{\partial t} + \vec{\nabla}.\left(n_{b0}\widetilde{\vec{v_b}}\right) + \vec{\nabla}.\left(\widetilde{n_b}\vec{v_{b0}}\right) = 0 \qquad (8)$$

The Gauss law becomes

$$\epsilon_0 \vec{\nabla}.\vec{E} = \rho = e(n_{b0} + \widetilde{n_b}) \qquad (9)$$

A wave-like solution is proposed for these perturbations hence the temporal and spatial derivatives are taken as

$$\frac{\partial}{\partial t} \rightarrow -i\omega \quad \text{and} \quad \nabla \rightarrow ik \qquad (10)$$

and by eliminating $\widetilde{n_b}$, $\widetilde{v_e}$ and $\tilde{E}$ from equations (7), (8) and (9), the dispersion relation for two-stream instability is obtained which is given as

$$\frac{\omega_{pe}^2}{\omega^2} + \frac{\omega_b^2}{(\omega - \vec{k}.\vec{v_b})^2} = 1 \qquad (11)$$

The dispersion relation for a plasma instability generated by a plasma beam interacting with a background plasma is given by [7]

$$1 = \omega_p^2 \left[\frac{m/M}{\omega^2} + \frac{1}{(\omega - k.v_0)^2}\right] \qquad (12)$$

However, if only one plasma species - electrons are considered then the ratio of electron to ion mass m/M = 1 and the above equation becomes [8]

$$\frac{\omega_p^2}{\omega^2} + \frac{\alpha \omega_p^2}{(\omega - \omega_d)^2} = 1 \qquad (13)$$

where

$$\alpha = \frac{\omega_b^2}{\omega_p^2} \quad \text{and} \quad \omega_d = k.v_{b0} \qquad (14)$$

This original dispersion relation is plotted in Figure 1.

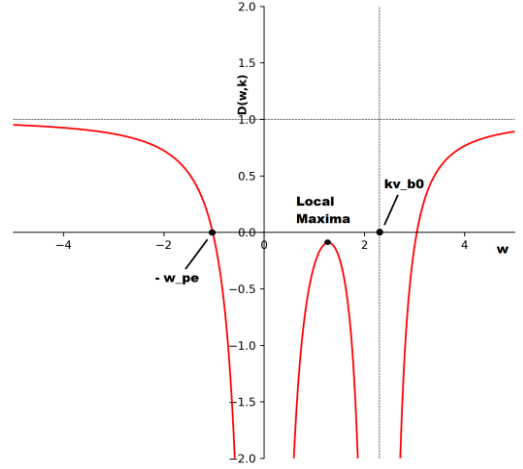

Figure 1. The original TSI dispersion relation.

## 3. Modified Dispersion Relation for LTSI

To modify the LTSI dispersion relation, the above analysis is extended which is as follows:
After linearizing, the electron momentum equation and continuity equation, for the electron beam system we get

$$\widetilde{n_b} = -\frac{i\vec{k}.\vec{E}}{(\omega - \vec{k}.\vec{v_b})^2}\left(\frac{n_{b0}e}{m_e}\right) \qquad (15)$$

For a plasma system, the momentum equation is given by

$$m_e n_{0e} \frac{\partial \vec{v_e}}{\partial t} = -en_{0e}\vec{E} - 3KT_e\vec{\nabla}\widetilde{n_e} \qquad (16)$$

and the continuity equation is given by

$$\frac{\partial \widetilde{n_e}}{\partial t} + n_{e0}\vec{\nabla}.\vec{v_e} = 0 \qquad (17)$$

and thereby replacing

$$\frac{\partial}{\partial t} \rightarrow -i\omega \quad \text{and} \quad \nabla \rightarrow ik$$

the momentum equation becomes

$$-m_e n_{e0} i\omega \vec{v_e} = -en_{e0}\vec{E} - 3KT_e i\vec{k}\widetilde{n_e} \qquad (18)$$

which gives $\vec{v_e}$ as

$$\vec{v_e} = \frac{e\vec{E}}{m_e i\omega} + \frac{3KT_e\vec{k}\widetilde{n_e}}{m_e n_{e0}\omega} \qquad (19)$$

The continuity equation becomes

$$-i\omega\widetilde{n_e} + n_{e0}i\vec{k}.\vec{v_e} = 0 \qquad (20)$$

which gives

$$\widetilde{n_e} = \frac{\vec{k}.\vec{v_e}}{\omega}n_{e0} \qquad (21)$$

Substituting $\vec{v_e}$ to $\widetilde{n_e}$

$$\widetilde{n_e} = \frac{\vec{k}.\left(\frac{e\vec{E}}{im_e\omega} + \frac{3KT_e\vec{k}\widetilde{n_e}}{m_e n_{e0}\omega}\right)}{\omega}n_{e0} \qquad (22)$$

Now,

$$v_{th}^2 = \frac{2KT_e}{m} \qquad (23)$$

hence, the above equation becomes

$$\widetilde{n_e} = \frac{\vec{k}}{\omega}\left(\frac{e\vec{E}}{im_e\omega} + \frac{3}{2}v_{th}^2\frac{\vec{k}\widetilde{n_e}}{n_{e0}\omega}\right)n_{e0} \qquad (24)$$

which can be written as

$$\widetilde{n_e} = \frac{(-i\vec{k}.\vec{E})\frac{n_{e0}e}{m_e\omega^2}}{1 - \frac{3}{2}v_{th}^2\frac{k^2}{\omega^2}} \qquad (25)$$

The beam-plasma system gives

$$\vec{\nabla}.\vec{E} = -\frac{\rho}{\epsilon_0}(\widetilde{n_e} + \widetilde{n_b}) + \frac{\rho_f}{\epsilon_0} \qquad (26)$$

Here ρ is the bound charge density and ρ$_f$ is free charge density

By substituting from equations (15) and (25), this can be written as

$$-i\vec{k}\cdot\vec{E} = -\frac{\rho}{\epsilon_0}\left[\frac{(-i\vec{k}\cdot\vec{E})\frac{n_{e0}e}{m_e\omega^2}}{1-\frac{3}{2}v_{th}^2\frac{k^2}{\omega^2}} + \frac{-i\vec{k}\cdot\vec{E}}{(\omega-\vec{k}\cdot\vec{v_{b0}})^2}\frac{n_{b0}e}{m_e}\right] + \frac{\rho_f}{\epsilon_0} \quad (27)$$

which gives

$$-i\vec{k}\cdot\vec{E}\left[1 - \frac{\frac{\omega_p^2}{\omega^2}}{1-\frac{3}{2}v_{th}^2\frac{k^2}{\omega^2}} + \frac{\omega_b^2}{(\omega-\vec{k}\cdot\vec{v_{b0}})^2}\right] = \frac{\rho_f}{\epsilon_0} \quad (28)$$

Now comparing with

$$\vec{\nabla}\cdot\vec{D} = \rho_f \quad \text{and} \quad \vec{\nabla}\cdot\epsilon_0\vec{E} = i\vec{k}\cdot\epsilon_0\vec{E} \quad (29)$$

By taking

$$i\vec{k}\cdot\epsilon\vec{E} = \rho_f \quad (30)$$

and we get

$$\varepsilon(k,\omega) = \epsilon_0\left[1 - \frac{\frac{\omega_p^2}{\omega^2}}{1-\frac{3}{2}v_{th}^2\frac{k^2}{\omega^2}} + \frac{\omega_b^2}{(\omega-\vec{k}\cdot\vec{v_{b0}})^2}\right] \quad (31)$$

For the normal mode of plasma oscillator

$$\varepsilon(k,\omega) = 0$$

The dispersion relation for this plasma system is

$$1 - \frac{\frac{\omega_p^2}{\omega^2}}{1-\frac{3}{2}v_{th}^2\frac{k^2}{\omega^2}} - \frac{\omega_b^2}{(\omega-\vec{k}\cdot\vec{v_{b0}})^2} = 0 \quad (32)$$

which can also be written as

$$1 - \frac{\omega_p^2}{\omega^2-(3/2)v_{th}^2 k^2} - \frac{\omega_b^2}{(\omega-\vec{k}\cdot\vec{v_{b0}})^2} = 0 \quad (33)$$

This is the two-stream instability dispersion relation modified with the inclusion of lunar hot electrons.

## 4. Modified LTSI Analysis

The modified LTSI dispersion relation as given in equation (33) can be written as

$$D(\omega,k) = 1 - \frac{\omega_p^2}{\omega^2-3/2 k^2 v_{th}^2} - \frac{\omega_b^2}{(\omega-kv_{b0})^2} = 0 \quad (34)$$

Now, the total electrons $n_e$ in the lunar ionosphere can be given as $n_e = n_{ec} + n_{eh}$ where, $n_{ec}$ are the cold (non-energetic) and $n_{eh}$ are the hot (energetic) lunar electrons.

The plasma oscillations generated by the lunar cold electrons is given as

$$\omega_{pc}^2 = \frac{n_{ec}e^2}{m_e\epsilon_0} \quad (35)$$

Assuming $n_{ec} = \beta n_e$, the angular plasma frequency for cold electrons becomes

$$\omega_{pc}^2 = \beta\left(\frac{n_{ec}e^2}{m_e\epsilon_0}\right) = \beta\omega_{pe}^2 \quad (36)$$

here $\omega_{pe}$ is the total angular plasma frequency in the lunar ionosphere due to $n_e$.

In a lunar plasma environment, the hot electron plasma frequency is given by

$$\omega_{ph}^2 = \frac{n_{eh}e^2}{m_e\epsilon_0} \quad (37)$$

where $\omega_{peh}$ is the angular hot electron frequency and $n_{eh}$ is the hot electron number density.

The dispersion relation of electron plasma waves is given by

$$\omega^2 = \omega_{pe}^2 + \frac{3}{2}k^2 v_{th}^2 \quad (38)$$

In this equation, the second term represents the electron perturbations due to its thermal velocities. Hence, the following approximation can be considered.

$$\omega_{ph}^2 \approx \frac{3}{2}k^2 v_{th}^2 \quad (39)$$

Similarly, for the lunar hot electrons

$$\omega_{ph}^2 = \gamma\omega_{pe}^2 \quad \text{and} \quad n_{eh} = \gamma n_e \quad (40)$$

Here, β and γ are in such a way that β + γ = 1. Also, we have

$$\alpha = \frac{n_b}{n_e} = \frac{\omega_b^2}{\omega_{pe}^2} \quad (41)$$

which gives

$$\omega_b^2 = \alpha\omega_{pe}^2$$

Inserting all these in equation (34), we get

$$D(\omega,k) = 1 - \frac{\beta\omega_p^2}{\omega^2-\gamma\omega_{pe}^2} - \frac{\alpha\omega_{pe}^2}{(\omega-\omega_d)^2} \quad (42)$$

This can be rearranged as

$$D(\omega,k) = 1 - \frac{\beta}{\left(\omega/\omega_{pe}\right)^2-\gamma} - \frac{\alpha}{\left(\omega/\omega_{pe}-\omega_d/\omega_{pe}\right)^2} \quad (43)$$

Taking $\omega/\omega_{pe} = X$ and $\omega_d/\omega_{pe} = X_0$

Equation (45) can be written as

$$D(\omega,k) = 1 - \frac{\beta}{X^2-\gamma} - \frac{\alpha}{(X-X_0)^2} \quad (44)$$

The nature of α, β and γ is defined as 0 < α, β, γ < 1 and γ < β.

The values of different variables are chosen as follows:
1. X = ω/$\omega_{pe}$ is taken between [0.5, 5]
2. $X_0$ = $\omega_d$/$\omega_{pe}$ = $kv_b$/$\omega_{pe}$ is taken as 1.5 (from reference 21); $v_b$ = 3.75 × 10$^5$ m/s (from [9]).
3. α = $n_b$/$n_e$, $n_b$ = 5 cm$^{-3}$ (typical solar wind electron density); $n_e$ = 10 - 300 cm$^{-3}$ (lunar electron number density) (from reference [9] and [10]). Hence, α is in the range [0.0167, 0.5].

Now, various fractions of hot lunar electrons is taken as

$$\frac{3}{2}k^2 v_{th}^2 = P(\%)\omega_p^2$$

Here, P (%) is the hot electron population which is 0.01 for 1%; 0.02 for 2%; 0.05 for 5%; 0.1 for 10%; 0.2 for 20% and 0.25 for 25%.

The dispersion relation with 1% to 25% hot lunar electrons is shown in Figure 1.

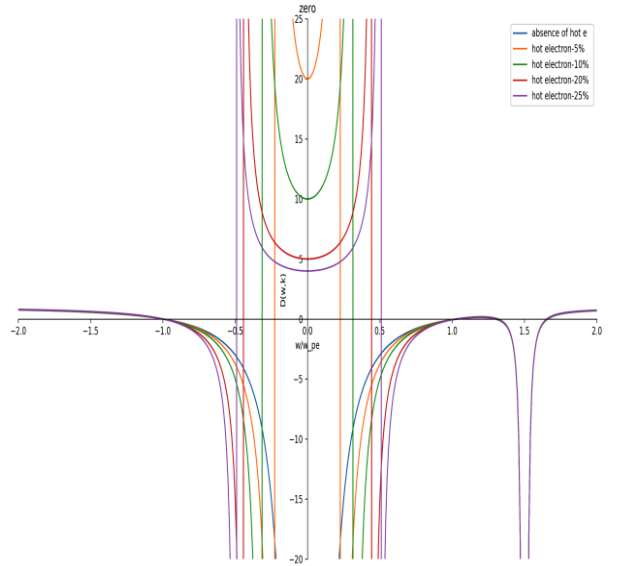

Figure 2. The modified dispersion relation plot with different fractions of hot lunar electrons.

As can be seen from Figure 1, for 1% hot lunar electrons the dispersion relation got changed and a new branch appears in the D(ω, k) axis. At the other places, the old and the new plots are superimposed over each other which indicate the overall stability of the plasma system. For 2% hot lunar electrons also the modified dispersion relation plot continues to change and the negative D(ω, k) axis starts emerging from the background indicating that the instability has started creeping into the plasma system. For 5% hot lunar electrons, the modified dispersion plot is becoming broader and the branches around the ω/$ω_{pe}$ axis emerged prominently on the positive D(ω, k) side indicating that the plasma system is becoming more unstable due to the growing instability. For 10% hot electrons, the modified dispersion relation plot got further expanded and the local maxima curve gets pushed to the positive D(ω, k) side indicating that the plasma system is now unstable. The distinct separation between the curves for pure cold electrons and the mixture of cold and hot electrons is also evident. For 20% hot electrons, the MTSI plot got expanded further with the curve specifically for hot electrons is coming down towards the 0 on the D(ω, k) axis making the plasma system highly unstable. It is also clear from Figure 1 that all the curves are in the positive quadrants of D(ω, k) axis. For 25% hot electron population, the plasma system becomes highly unstable as all the indicators are now prominent. The hot electron component curve of the modified dispersion relation of the lunar TSI is at its minimum and there is a maximum deviation between the pure cold and the combination of cold and hot electrons in the negative D(ω, k) quadrants. All other curves are in the positive quadrants of D(ω, k) axis showing the instability.

Generally, for an energetic electron population of around 5-10%, a plasma system is capable of becoming unstable even in the absence of an external trigger. In this analysis, a hot electron fraction of 20% and 25% is taken to consider the solar extreme transient events which can lead to the heating of lunar electrons to enhancing the amount of such proportions of hot electrons.

## 5. Concluding Remarks

The solar wind electrons interact with the lunar electrons which has both the non-energetic (cold) and energetic (hot) electron components. The two-stream instability generation study is extended by including the hot lunar electrons in proportion with the cold component of lunar plasma. The dispersion relation of the TSI is modified and the effect of including various hot electron populations such as 1%, 2%, 5%, 10%, 20% and 25% of the total lunar electrons on the TSI dispersion relation is analyzed. It is observed that the plasma system becomes more and more unstable as the fraction of hot lunar electrons is increased. The deviation in the dispersion relation curves of the lunar TSI between the purely cold and a mixture of cold and hot electrons becomes evident with an increase in the hot lunar electron population.

## 6. References

1. M. Chakraborty, Vipin K. Yadav and Rajneesh Kumar "Streaming Instability Generation in Lunar Plasma Environment", *Proc. of URSI-RCRS-2022*; December 1-4, 2022; IIT Indore; pp. 509-512, doi:10.23919/URSI-RCRS56822.2022.10118524.

2. M. Chakraborty, Vipin K. Yadav and Rajneesh Kumar "Two Stream Instability Generation in the Lunar ionosphere", *Advances in Space Research*, **71**, 6, 2023, pp. 2954-2966, doi:10.1016/j.asr.2022.11.050

3. B. E. Goldstein, "Observations of electrons at the lunar surface," *Journal of Geophysical Research*, **79**, 1, pp. 23-35, 1974, doi:10.1029/JA079i001p00023.

4. Y. Harada, S. Machida, J. S. Halekas, A. R. Poppe, and J. P. McFadden, "Artemis observations of lunar dayside plasma in the terrestrial magnetotail lobe", *Journal of Geophysical Research: Space Physics*, **118**, 6, pp. 3042-3054, 2013, 10.1002/jgra.50296.

5. J. T. Gosling, "The Solar Wind," in *Encyclopedia of the Solar System*, Eds. L. A. A. McFadden, P. R. Weissman, and T. V. Johnson, pp. 99-116, 2007.

6. J. S. Halekas, G. T. Delory, R. P. Lin, T. J. Stubbs, and W. M. Farrell, "Lunar prospector observations of the electrostatic potential of the lunar surface and its response to incident currents," *Journal of Geophysical Research: Space Physics*, **113**, A09, pp. A09102, 2008, doi:10.1029/2008JA013194.

7. F. F. Chen, "Introduction to plasma physics and controlled fusion," (Springer, Springer, New York, 2016).

8. D. Anderson, R. Fedele, and M. Lisak, "A tutorial presentation of the two stream instability and landau damping," *American Journal of Physics*, **69**, 12, pp. 1262-1266, 2001, doi:10.1119/1.1407252.

9. A. Piel, "Plasma Physics: An Introduction to Laboratory, Space, and Fusion Plasmas", Chapter 8 "Instabilities, 2010, Springer-Verlag Berlin, Heidelberg, doi:10.1007/978-3-642-10491-6

10. R. K. Choudhary, et al., "On the origin of the ionosphere at the moon using results from chandrayaan-1 s band radio occultation experiment and a photochemical model", *Geophysical Research Letters*, **43**, 19, pp. 10025 - 10033, 2016, 10025, doi:10.1002/2016GL070612.